\documentclass[11pt]{article}

\usepackage{bm}
\usepackage[normalem]{ulem}
\usepackage[utf8]{inputenc}
\usepackage[T1]{fontenc}
\usepackage{mathptmx}
\usepackage{etoolbox}
\usepackage{xcolor}
\usepackage{physics}

\usepackage[a4paper, margin=0.7in]{geometry}
\usepackage{authblk}
\usepackage{graphicx}
\usepackage{dcolumn}
\usepackage{amsmath}
\usepackage{amssymb}
\usepackage{textcomp}
\usepackage{verbatim}
\usepackage{float}
\usepackage[symbol]{footmisc}
\setfnsymbol{wiley}
\usepackage{hyperref}
\hypersetup{
    colorlinks=true,
    linkcolor=blue,
    filecolor=blue,
    urlcolor=blue,
    citecolor = blue,
    pdfauthor=author,
}

\usepackage{multirow}
\usepackage[T1]{fontenc}
\usepackage[normalem]{ulem}
\usepackage[numbers,sort&compress]{natbib}

\begin{document}

\title{Elliptical micropillars for efficient generation and detection of coherent acoustic phonons}
% Force line breaks with \\
\author[1]{C. Xiang}
\author[1]{A. Rodriguez}
\author[1]{E. Cardozo de Oliveira}
\author[1]{L. Le Gratiet}
\author[1]{I. Sagnes}
\author[1]{M. Morassi}
\author[1]{A. Lemaître} 
\author[1]{N.D. Lanzillotti-Kimura\footnote{daniel.kimura@c2n.upsaclay.fr}}
\affil[1]{Université Paris-Saclay, CNRS, Centre de Nanosciences et de Nanotechnologies, 91120 Palaiseau, France}

\date{}
\maketitle

\begin{abstract}

\textbf{ABSTRACT.} Coherent acoustic phonon generation and detection assisted by optical resonances are at the core of efficient optophononic transduction processes. However, when dealing with a single optical resonance, the optimum generation and detection conditions take place at different laser wavelengths, i.e. different detunings from the cavity mode. In this work, we theoretically propose and experimentally demonstrate the use of elliptical micropillars to reach these conditions simultaneously at a single wavelength. Elliptical micropillar optophononic resonators present two optical modes with orthogonal polarizations at different wavelengths. By employing a cross-polarized scheme pump-probe experiment, we exploit the mode splitting and couple the pump beam to one mode while the probe is detuned from the other one. In this way, at a particular micropillar ellipticity, both phonon generation and detection processes are enhanced. We report an enhancement of a factor of $\sim$3.1 when comparing the signals from elliptical and circular micropillars.  Our findings constitute a step forward in tailoring the light-matter interaction for more efficient ultrahigh-frequency optophononic devices.
\end{abstract}

\maketitle

\section{Introduction}

The efficient manipulation of gigahertz (GHz) acoustic phonons at the nanoscale is crucial for transforming various technological applications, from quantum information to data processing.~\cite{priya_perspectives_2023,balandin_nanophononics_2005,volz_nanophononics_2016} For instance, the long phonon lifetimes and short wavelengths make them good candidates for information carriers and quantum memories. The coupling of gigahertz acoustic phonons with superconducting qubits~\cite{oconnell_quantum_2010,chu_quantum_2017,von_lupke_parity_2022} or cavity exciton-polaritons,~\cite{chafatinos_polariton-driven_2020,kuznetsov_microcavity_2023} using GHz bulk acoustic wave (BAW) resonators, exemplifies the potential of the field. Optophononic resonators based on GaAs/AlAs semiconductor microcavities, on the other hand, offer versatile platforms reaching frequencies up to a few terahertz.~\cite{Fainstein2007,lanzillotti-kimura_coherent_2007,lanzillotti-kimura_phonon_2007} The resonant frequency of acoustic phonons in the vertical direction depends on layer thicknesses, controlled through advanced techniques like molecular beam epitaxy.~\cite{lanzillotti-kimura_acoustic_2007} Tailoring the optophononic properties of these resonators by defining 3D structures, such as micropillars, enhances light-matter interactions due to the lateral confinement that localizes the optical and acoustic fields.~\cite{anguiano_micropillar_2017,lamberti_optomechanical_2017,esmann_brillouin_2019-1} Additionally, these systems allow the integration of acoustic resonators with other solid-state platforms, such as exciton-polaritons or single photon sources, when defining quantum wells or quantum dots at the cavity spacer.~\cite{bajoni_polariton_2008,carlon_zambon_enhanced_2022,senellart_high-performance_2017} However, efficient transduction for optimum phonon generation and detection at the nanoscale remains a major challenge. This issue has been explored in acoustoplasmonics \cite{obrien_ultrafast_2014,lanzillotti-kimura_polarization-controlled_2018,baidaOpticalDetectionFemtosecond2011,baidaUltrafastNonlinearOptical2011,berteAcousticFarFieldHypersonic2018,pobletAcousticCouplingPlasmonic2021}, and recently, it has been proposed the use of circularly polarized light with plasmonic structures holding chiral geometry to selectively enhance phonon generation and detection.~\cite{castillo_lopez_de_larrinzar_towards_2023}
In this work, we present an alternative strategy based on semiconductor micropillars that simultaneously confine NIR light and acoustic phonons. 

Manipulating ultrahigh frequency acoustic phonons usually requires all-optical experiments, such as time-domain Brillouin scattering.~\cite{thomsen_coherent_1984,thomsen_surface_1986,ruello_physical_2015} In microcavity optophononic resonators the roadblock arises from the fact that the phonon generation process is most efficient when the excitation laser matches the optical mode resonance, while phonon detection, linked to optical reflectivity changes due to changes in the refractive index, is most sensitive at the slope of the optical mode.~\cite{lanzillotti-kimura_enhanced_2011,lanzillotti-kimura_theory_2011,anguiano_optical_2018} This limitation can be overcome by using two lasers slightly detuned in order to reach the optimal condition.~\cite{baidaUltrafastNonlinearOptical2011,peronneTwocolorFemtosecondStrobe2013} While an effective and straightforward strategy, this solution requires two laser oscillators. Another approach is to change the angle of incidence of the excitation laser so that the in-plane component of the optical dispersion relation leads to a different resonant mode.~\cite{lanzillotti-kimura_enhanced_2011} Although successful in DBR-based planar microcavities, this technique is not feasible with micropillars due to lateral confinement constraints. To date, in the case of micropillars, no phonon enhancement strategies have been proposed, except the integration of pillar microcavities with single-mode fibers, enabling long-term stabilization and reproducible plug-and-play experiments.~\cite{ortiz_fiber-integrated_2020}

In this scenario, elliptical micropillars emerge as promising options for enhancing device efficiency. Their elliptical shape induces birefringence, causing the confined optical mode to split into two non-degenerate polarized modes aligned with the micropillar axes.~\cite{gayral_optical_1998,whittaker_high_2007,sebald_optical_2011} The size and ellipticity of the micropillars determine the mode splitting. Elliptical micropillars have been employed to improve the emission of single photon sources based on quantum dots embedded in microcavity spacer.~\cite{wang_towards_2019,daraei_control_2006,reitzenstein_polarization-dependent_2010,gerhardt_polarization-dependent_2019} The micropillar ellipticities have been recently explored to manipulate the polarization selection rules of Brillouin scattering.~\cite{rodriguez_brillouin_2023} 

Here, we theoretically propose and experimentally demonstrate how to enhance phonon generation and detection simultaneously using elliptical micropillars. Through experimentation with micropillars of different ellipticities, we establish that an optimal ellipticity significantly amplifies the phonon generation-detection process.

\section{Methods}

\subsection{Experimental details}

\begin{figure*}
\center
    \includegraphics[scale=1]{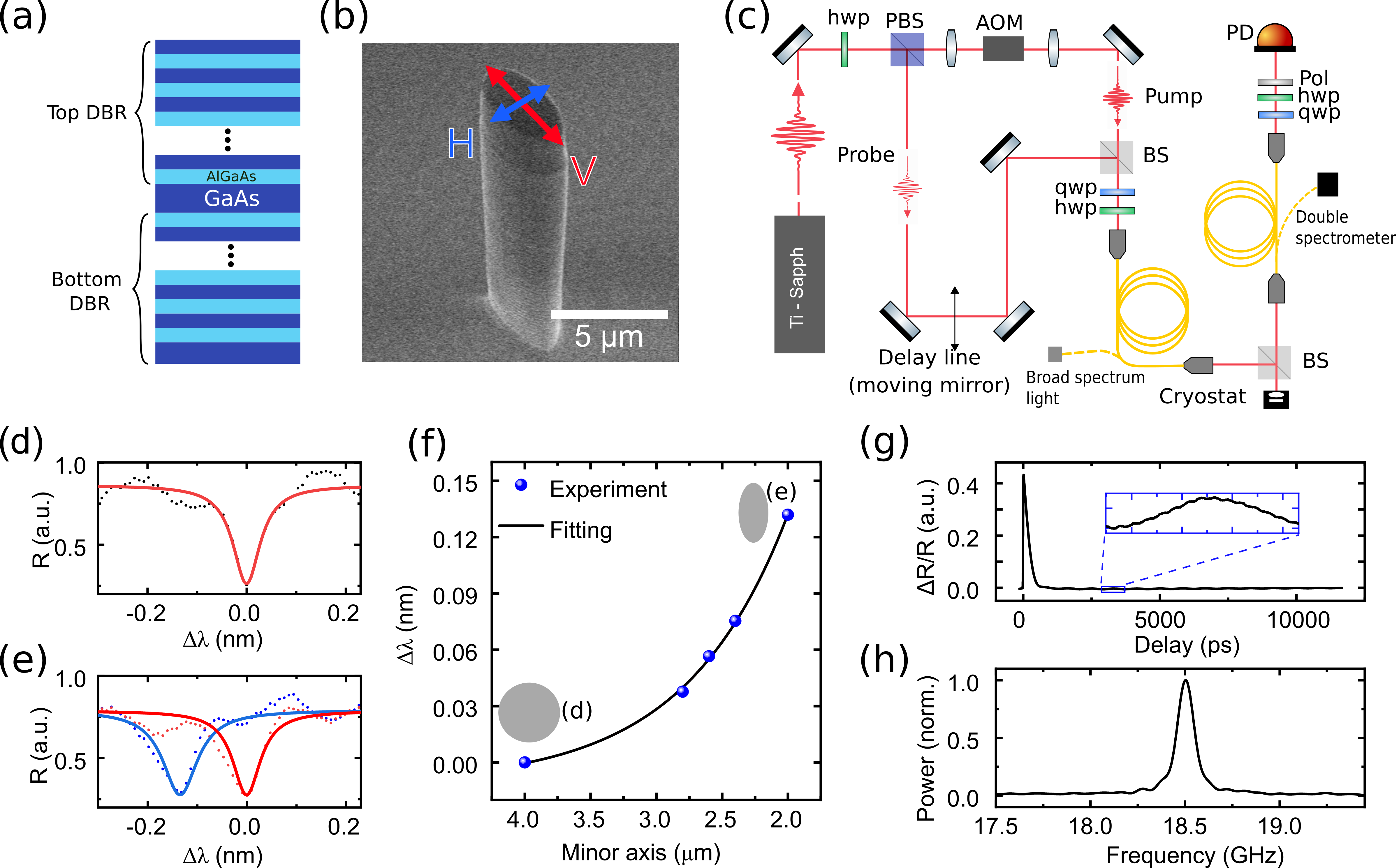}
    \caption{(a) Schematic of the vertical structure of the micropillars. Dark (light) blue corresponds to GaAs (AlAs). (b) SEM image of an elliptical micropillar. The red (blue) arrow indicates the major (minor) axis of the cross-section. (c) Schematic of pump-probe experimental setup. \textit{qwp}, \textit{hwp}, BS, PBS, Pol, AOM, and PD stand for quarter (half) waveplate, beam splitter, polarization beam splitter, polarizer, acousto-optical modulator, and photodiode, respectively. (d) Optical reflectivity of a $4-\mu$m-diameter circular micropillar. (e) Optical reflectivities of an elliptical micropillar with major (red) and minor (blue) axes of 4~$\mu$m and 2~$\mu$m, respectively. (f) Optical mode splitting of five micropillars with the minor axis of 4, 2.8, 2.6, 2.4, and 2~$\mu$m, while the major axis is 4~$\mu$m.  (g) Differential reflectivity in the time domain. Inset: zoom-in of the reflectivity time trace, showing low amplitude oscillations due to coherent acoustic phonons. (h) Fast Fourier transform of the time trace shown in (g).}
    \label{setup}
\end{figure*}

The studied structures are based on an optophononic microcavity grown on a (001)-oriented GaAs substrate by molecular-beam epitaxy. It consists of two distributed Bragg reflectors (DBRs) enclosing a $\lambda$/2 GaAs spacer (Fig.~\ref{setup}(a)).~\cite{rodriguez_brillouin_2023,bajoni_polariton_2008} The top (bottom) DBR is formed by 25 (29) periods of Ga$_{0.9}$Al$_{0.1}$As/Ga$_{0.05}$Al$_{0.95}$As ($\lambda$/4,$\lambda$/4). This design constitutes an optophononic resonator confining near-infrared photons at around $\lambda$ = 900~nm and acoustic phonons at $\sim$18~GHz,~\cite{anguiano_micropillar_2017,esmann_brillouin_2019-1,ortiz_topological_2021} with Q-factors of $\sim$11000. Elliptical micropillars are processed from the planar cavity by electron beam lithography and inductively coupled plasma etching. Figure~\ref{setup}(b) shows the scanning electron microscope (SEM) image of a micropillar with an elliptical cross-section. Here, we define the major axis as $a$ and the minor axis as $b$.

The schematics of the time-domain Brillouin scattering technique,~\cite{thomsen_coherent_1984,thomsen_surface_1986} used to study coherent phonon generation and detection, is shown in Fig.~\ref{setup}(c). Picosecond pulses from a Ti-Sapph laser are split into two linearly polarized beams with orthogonal polarizations (pump and probe) by a polarizing beamsplitter. The pump generates phonons via the deformation potential mechanism.~\cite{ruello_physical_2015} The presence of coherent acoustic phonons in the structure modulates the material's refractive index. The probe beam then detects transient reflectivity changes. By scanning the time delay between the pump and probe beams with a mechanical delay line we can reconstruct the reflectivity time trace of the structure. Both beams merge at the beam splitter and are coupled to a single-mode fiber. They are focused on the sample to a spot of $\sim$2.2~$\mu$m diameter with an objective lens of 0.7 NA. The reflected light is coupled to a different fiber and sent to a fast photodetector (PD). A combination of quarter (\textit{qwp}) and half (\textit{hwp}) waveplates is used to manipulate the polarization of the two beams. For the collection, the pump is filtered out by a cross-polarization filtering scheme with a second set of \textit{qwp} and \textit{hwp} and polarizer before the PD. The signal is then analyzed by a lock-in amplifier. All the measurements are performed at room temperature. To observe the optical modes of the elliptical pillars, reflectivity measurements can be performed in the same setup.~\cite{rodriguez_brillouin_2023} A broadband light is adapted to the incident path, and the reflected light is sent to a double spectrometer and detected by a charge-coupled device. A removable polarizer is attached to the collection path, before the fiber, to select the polarization of interest.

The reflectivity spectra of micropillars with a circular ($a$ = $b$ = 4~$\mu$m) and an elliptical (a = 4~$\mu$m and b = 2~$\mu$m) cross-section are shown in Figs.~\ref{setup}(d) and ~\ref{setup}(e), respectively. The reflectivity of the circular micropillar reveals a cavity mode centered at 902.27~nm, with a full width at half maximum of $\sim$0.08~nm. Only one optical mode is present and it has no preferential polarization axis. The eigenfrequency is inversely proportional to the micropillar radius. In the case of the elliptical micropillar, the degeneracy of the eigenfrequency is lifted and we can observe two optical cavity modes in the reflectivity according to each axis of the elliptical cross-section and are conventionally associated with the V and H polarization for the major ($a$) and minor ($b$) axis, respectively. The V and H modes are at 900.90~nm and 900.77~nm, respectively, exhibiting a splitting of $\Delta\lambda=0.13$~nm.

The optical mode splitting dependence with the minor axis $b$ (while the major axis $a$ remains constant at $4~\mu m$) is displayed in Fig.~\ref{setup}(f). The splitting increases by decreasing $b$. In the pump-probe experiment, we exploit the cross-polarization of the beams combined with the mode splitting of the elliptical pillars. The polarization of the pump beam is aligned with the major axis of the micropillar, and therefore it is coupled to the V mode of the elliptical pillar. Accordingly, the probe is automatically parallel to the mode along the minor axis (H mode). A typical time-dependent differential reflectivity is displayed in Fig.~\ref{setup}(g). A rapid change and exponential decay within $\sim500$~ps is present. After removing the low-frequency component and performing a Fourier transform, we observe a peak at around 18.5 GHz in the spectrum (Fig.~\ref{setup}(h)) corresponding to the fundamental mode of the micropillar. 

\subsection{Model implementation}

\begin{figure}[h]
\center
    \includegraphics[scale=1]{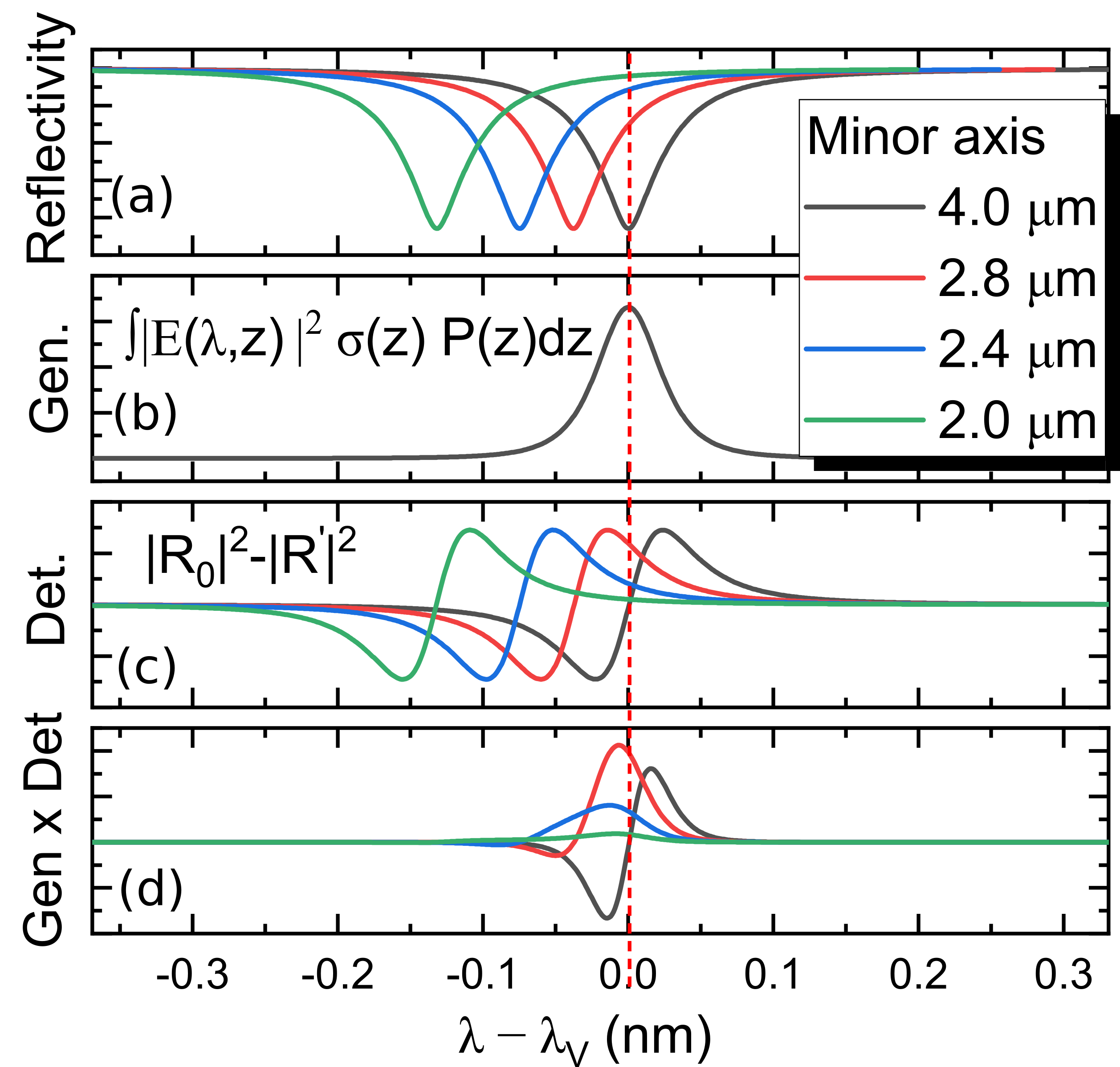}
    \caption{Simulation of phonon generation and detection in four elliptical micropillars with minor axis $b=$ 4~$\mu$m (black), 2.8~$\mu$m (red), 2.4~$\mu$m (blue) and 2~$\mu$m (green), whereas the major axis $a=4~\mu$m is fixed. (a) Simulated reflectivity of 4 pillars as a function of optical mode shifting relative to consistent V mode. (b) and (c) correspond to the phonon generation and detection spectra, respectively. (d) Product of the generation and detection spectrum.}
    \label{simulation}
\end{figure}

We compute the optical reflectivities and the optical and strain fields using the standard transfer matrix method for multilayered structures, considering the appropriate parameters for electromagnetic -- refractive index ($n$) and speed of light ($c$)-- and acoustic waves -- mass density ($\rho$) and speed of sound ($v$).~\cite{matsuda_reflection_2002,lanzillotti-kimura_theory_2011,Fainstein2007,lanzillotti-kimura_phonon_2007} To simulate the optical mode splitting of elliptical micropillars we assume two different structures (corresponding to each axis) for each ellipticity: first, we calculate the reflectivity of the V mode (associated with the major axis $a = 4~\mu m$) at the wavelength $\lambda_{V}$; and for the H mode, we tune the overall refractive index of the multilayered structure so that the optical resonance is shifted, and the wavelength difference between the H and V modes match the experimental splitting. The reflectivity of the mode along the minor axis for four different cases (4, 2.8, 2.4, and 2~$\mu m$) are shown in Fig.~\ref{simulation}(a). Note that the reflectivity spectrum for the case $b=a=4~\mu m$ falls in the circular micropillar situation, in which the mode is independent of the axis orientation.

The amplitude of the coherent phonon generation $g(\omega,\lambda)$ is given by the overlap integral~\cite{lanzillotti-kimura_theory_2011,Fainstein2007}
\begin{equation}
g(\omega,\lambda)=\int p(z) \eta(\omega, z)|E(\lambda, z)|^2 dz,
\end{equation}
where $\omega=2\pi f$, $\lambda$, $z$, $P(z)$, $\eta(\omega, z)$ and $E(\lambda, z)$ correspond to the acoustic phonon angular frequency, the optical wavelength, the position along the multilayer stacking axis, the material dependent photoelastic constant, the acoustic strain, and the laser electric field, respectively. The results of the phonon generation amplitude as a function of the excitation laser wavelength at the acoustic resonant frequency $f = 18.5$~GHz are displayed in Fig.~\ref{simulation}(b). The peak amplitude centered at $\lambda_{V}$ is a consequence of the maximum overlap integral between electric and strain fields.

The detection efficiency, independent of the generation process, can be calculated assuming a photoelastic interaction. The acoustic phonons present in the structure modify the refractive indices which leads to a differential modification of the optical reflectivity. We consider a perturbative term ($\Delta n$) for the refractive index proportional to the strain field at $f=18.5$~GHz and to the photoelastic constant $P(z)$. We then calculate the perturbed reflectivity $R$ with the modified strain $n'=n_{0}+\Delta n$. Finally, the variation in reflectivity is calculated by~\cite{lanzillotti-kimura_theory_2011}
\begin{equation}
\Delta R(\lambda, \omega)=\left|R_0(\lambda)\right|^2-\left|R(\lambda, \omega)\right|^2
\end{equation}
where $R_{0}$ is the unperturbed reflectivity. The detection process is performed by the probe beam which is coupled to the H mode, therefore these calculations are repeated every minor axis, as shown in Fig.~\ref{simulation}(c). One can see that the detection efficiency is always zero at the center of the optical cavity mode, and maximum (with opposite signs) at both slopes. 

The product of the phonon generation and detection spectra, shown in Fig.~\ref{simulation}(d), maps the resulting wavelength-dependent phonon amplitude efficiency. However, the pulsed laser employed in the experiment has a spectral Gaussian profile with an associated FWHM of around twice the optical cavity linewidth. Each wavelength component of the laser is then susceptible to a different phonon generation-detection efficiency. To account for that, we calculate the product of a Gaussian centered at $\lambda_{V}$, related to the laser spectral profile, with the function displayed in Fig.~\ref{simulation}(d). The simulated phonon amplitude is then a result of an overlap integral between the laser spectrum and the generation and detection efficiencies. In a circular micropillar, one can predict that the phonon amplitude will be close to zero due to the opposite signs of the detection efficiency, which can be comprehended as counter-phase reflectivity modulation that cancels the signal. In addition, the case with a minor axis of 2.8~$\mu$m leads to the strongest signal, and finally, the extreme case of 2~$\mu$m leads to a weak phonon signal again, as the H mode is well detuned from the laser. 

\section{Results and discussion}

\begin{figure}
\center
    \includegraphics[scale=0.65]{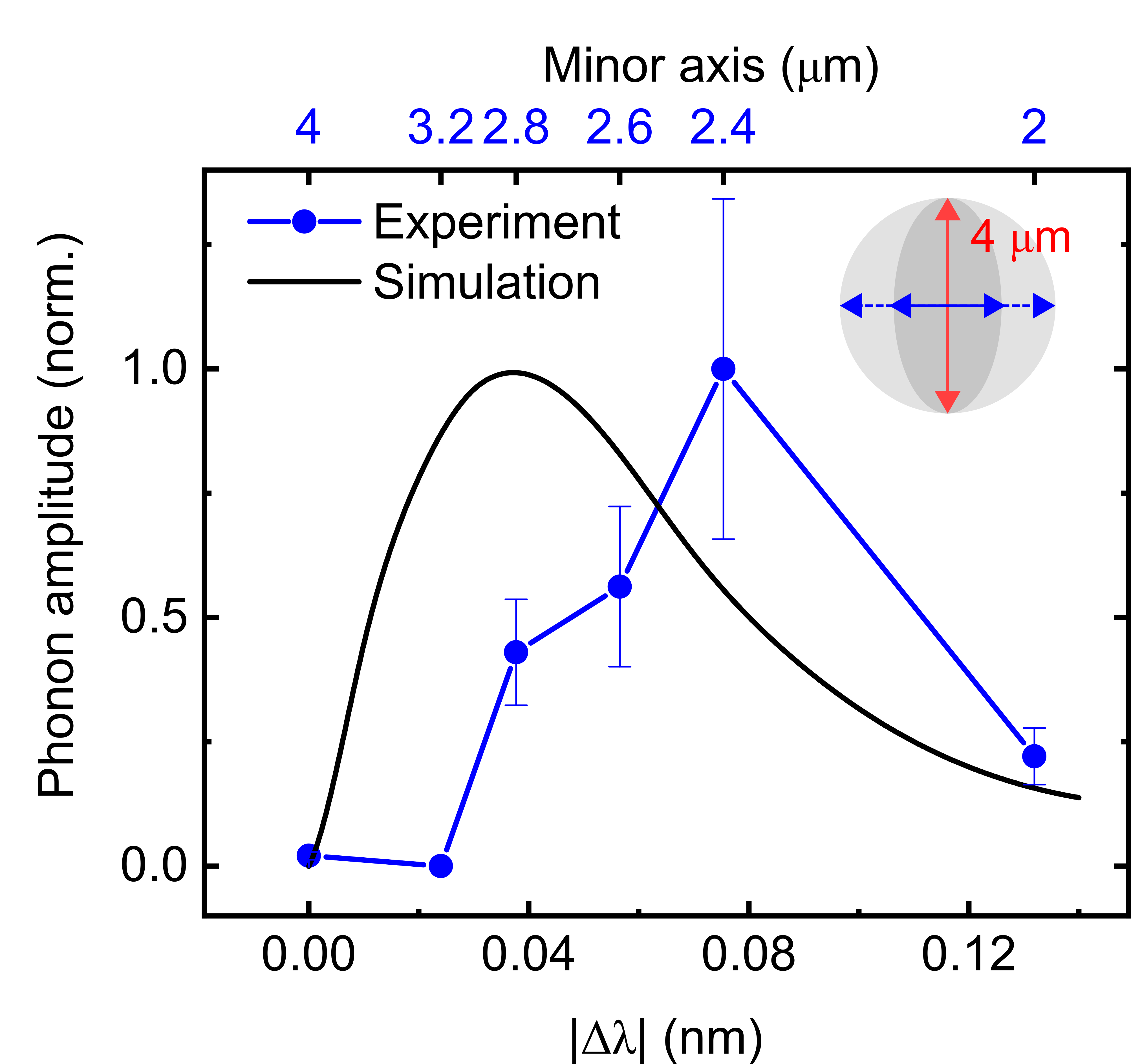}
    \caption{The amplitude of the phonon signal as a function of the minor axis of the pillars (top axis) and the mode splitting (bottom axis). The experimental and simulation results are presented in blue and black, respectively. The error bars correspond to the fluctuation of multiple experiments.}
    \label{fig:five_pillar}
\end{figure}

Figure~\ref{fig:five_pillar} presents the normalized pump-probe phonon amplitude as a function of the splitting between the H and V optical modes. The case of zero splitting corresponds to a circular micropillar with 4~$\mu$m-diameter. The error bars are associated with the divergence of the amplitude among five repeated measurements on each pillar. The experiment is performed with pump and probe powers of 40~$\mu$W and 100~$\mu$W, respectively. The laser is tuned in resonance with the V mode so that the phonon generation efficiency is constant.

For circular micropillars, the phonon amplitude is nearly zero. In this case, the counter-phase reflectivity modulation at both slopes of the cavity is equally sensed by the probe, which cancels the signal. By increasing the ellipticity, the phonon detection increases and reaches a maximum amplitude for the pillar with b = 2.4~$\mu$m. At the other extreme, the splitting of the modes leads to a weakly coupled probe into the H mode, and thus, the detected phonon signal decreases. The linewidths of the laser pulse and the optical cavity mode are comparable. This leads to a strongly wavelength-dependent coherent phonon generation-detection process for the highest phonon amplitude case. The combination of small laser wavelength fluctuations with slightly different mode matching in each experiment increases imprecision. Despite the large error bars, the phonon amplitude within the experimental error is still higher than the amplitudes at adjacent ellipticities.

Although the trend is compatible with the predicted one, with a maximum phonon amplitude in between the two extreme cases, there is a clear offset between the experimental results and the simulation curve shown in Figure~\ref{fig:five_pillar}, according to the model described in the previous section. The computed values show a maximum phonon amplitude for an elliptical micropillar with $b=2.8~\mu$m, with a mode splitting of $\sim0.038$~nm, i.e., $\sim0.032$~nm less than the measured one. We associate this discrepancy to the warming up of the sample due to the  relatively high power of both pump and probe beams. 

\begin{figure}
\center
    \includegraphics[scale=0.5]{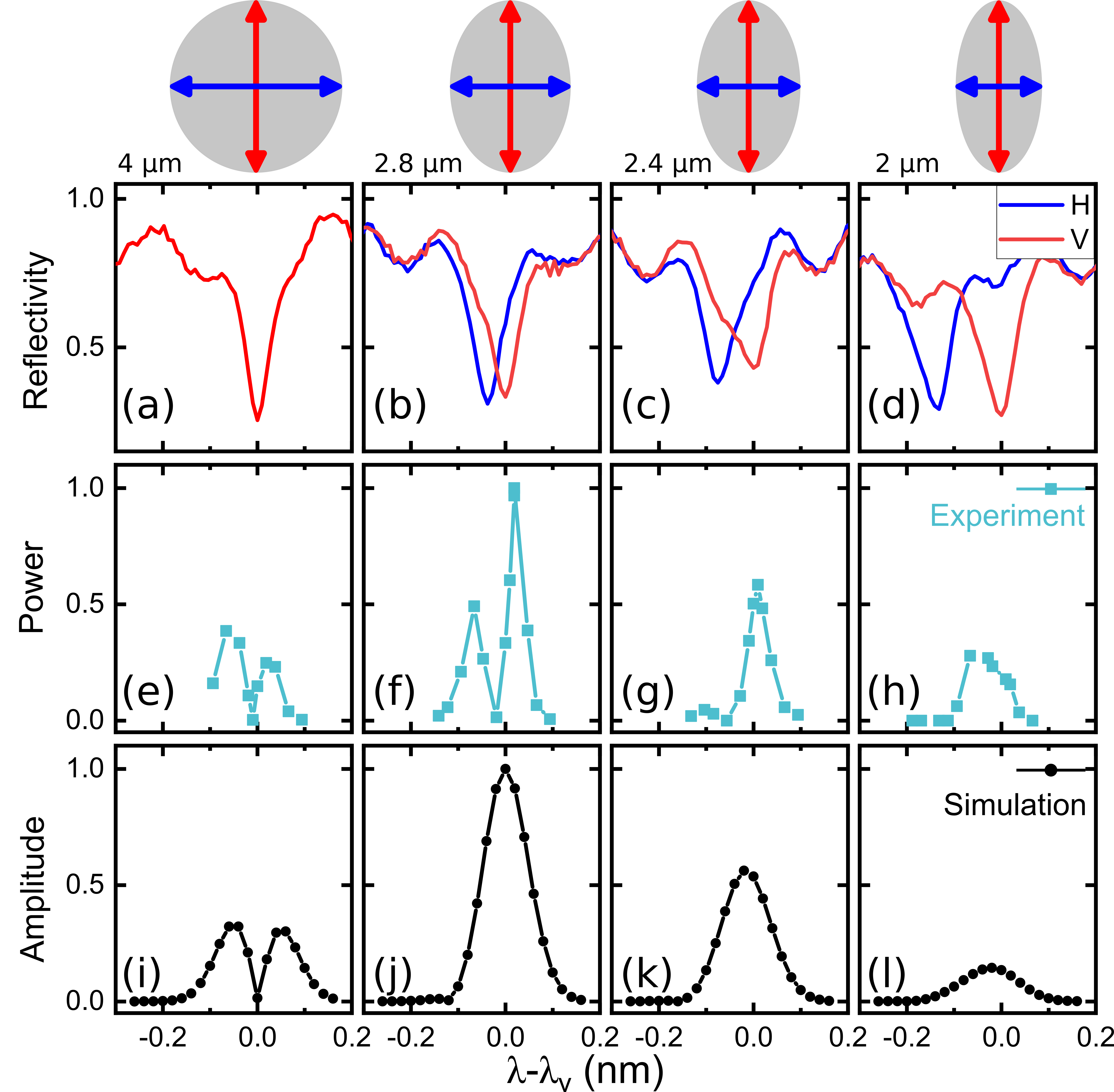}
    \caption{ Wavelength dependent phonon signal on micropillars. Each column corresponds to one micropillar, as shown on top of the figure. (a)-(d) Experimental reflectivity. (e)-(h) Laser wavelength dependence of the power of phonon signal. (i)-(l) are the simulated phonon amplitude as a function of laser detuning.}
    \label{wl_scan}
\end{figure}

We perform pump-probe experiments as a function of the laser wavelength for four elliptical pillars across the optical cavity modes in steps of $\sim 0.01$~nm. The pump and probe powers are reduced to 30~$\mu$W and 50~$\mu$W, respectively, corresponding to respective reductions of 25$\%$ and 50$\%$, compared to the previous experiment. For reference, the reflectivity and cross-section of the micropillars are displayed in Figs.~\ref{wl_scan}(a)-\ref{wl_scan}(d). In Figs.~\ref{wl_scan}(e)-\ref{wl_scan}(h), we present the measured phonon intensity as a function of the laser wavelength. We simulate the laser detuning by shifting the center wavelength of the Gaussian profile that is multiplied by the phonon generation-detection function. In Figs.~\ref{wl_scan}(i)-\ref{wl_scan}(l), the plots display the phonon amplitude, i.e., the integral of the resulting product, as a function of the laser detuning.

For the circular micropillar (Fig.~\ref{wl_scan}(e)), when there is no symmetry breaking in the cross-section, the results are aligned with previously documented results~\cite{lanzillotti-kimura_theory_2011,anguiano_optical_2018}. The highest amplitude occurs along the slopes of the optical cavity, indicating that the detection process is directly associated with the modulation of the refractive index of the structure. These results are consistent with the simulations depicted in Fig.~\ref{wl_scan}(i). By introducing the ellipticity, the shape of the wavelength-dependent phonon amplitude changes. Notably, the intensity now reaches the maximum when the minor axis is 2.8~$\mu$m (Fig.~\ref{wl_scan}(f)), as previously expected, which indicates that warming-up effects are less pronounced. Note that for this specific elliptical cross-section, the phonon signal is strongly dependent on the laser wavelength close to the maximum amplitude. A wavelength shift in the response is observed, which might still be associated with a warming up of the sample. Discrepancies might stem from unaccounted factors like fabrication imperfections, causing incomplete decoupling of optical cavities, or laser disturbances leading to slight shifts in optical resonances. Additional simulations exploring finer changes in ellipticity are detailed in the Appendix A. Nevertheless, we observe an enhancement of a factor of $\sim$2.6 when comparing the maximum experimental amplitude of Fig.~\ref{wl_scan}(f) with the circular micropillar (Fig.~\ref{wl_scan}(e)), and theoretically, this factor reaches $\sim$3.1. Finally, for the elliptical pillars with $b=2.4~\mu$m (Fig.~\ref{wl_scan}(g)) and $b=2~\mu$m (Fig.~\ref{wl_scan}(h)), we observe one distinct peak, which matches well with the simulation (Figs.~\ref{wl_scan}(k) and \ref{wl_scan}(i)). 

These results demonstrate the feasibility of engineering more efficient transduction of acoustic phonons by shaping the elliptical cross-section of micropillars for practical applications.

\section{Conclusions}

In this study, we have investigated the unique properties of optophononic micropillar cavities with elliptical cross-sections, focusing on enhancing both coherent phonon generation and detection processes simultaneously. Our findings reveal a significant influence of micropillar ellipticity on phonon signals, demonstrating maximum amplitude at specific major and minor axis lengths. When comparing the signals from elliptical and circular micropillars, we report an enhancement of a factor of $\sim$3.1. This study marks a significant advancement in the development of an efficient ultrahigh frequency acoustic phonon platform, crucial for practical applications in areas such as quantum technologies and data processing.
  
\section*{Acknowledgements}
The authors gratefully acknowledge P. Priya for fruitful discussions and support at an early stage of the project. The authors acknowledge funding from European Research Council Consolidator Grant No.101045089 (T-Recs). This work was supported by the European Commission in the form of the H2020 FET Proactive project No. 824140 (TOCHA), the French RENATECH network, and through a public grant overseen by the ANR as part of the “Investissements d'Avenir” Program (Labex NanoSaclay Grant No. ANR-10-LABX-0035).

\section*{Appendix A}

\begin{figure*}[ht]
\center
    \includegraphics[scale=0.5]{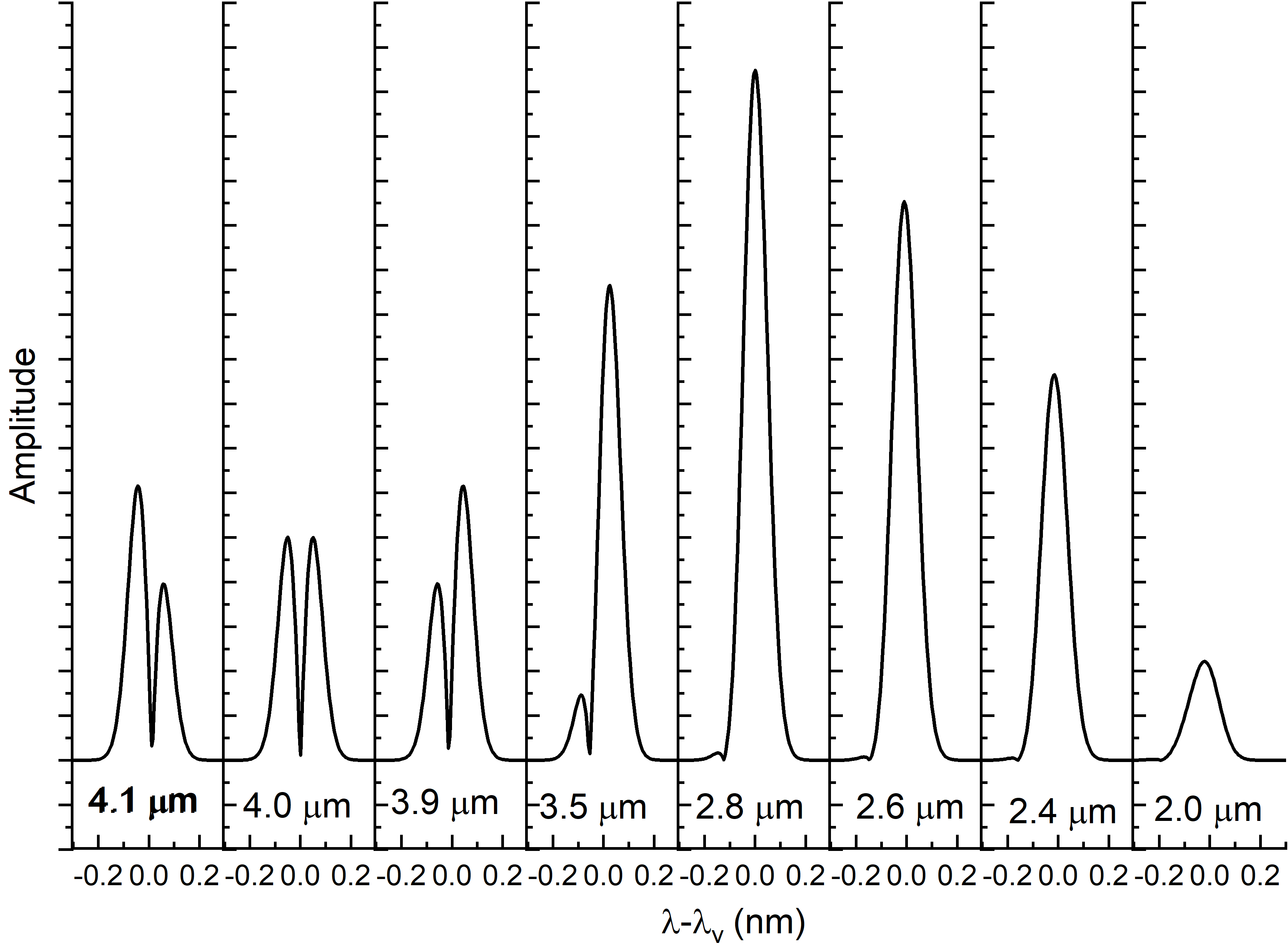}
    \caption{ Simulated wavelength-dependent phonon signal on micropillars with different ellipticities.}
    \label{Appendix}
\end{figure*}

Figure~\ref{Appendix} displays the simulated phonon amplitude as a function of the laser wavelength detuning for different micropillars. This shows that small changes in the ellipticity can strongly affect the shape and amplitude of the signal.

\end{document}